\documentclass{INTERSPEECH2023}
\usepackage{amsmath}
\usepackage{graphicx}
\usepackage{array}
\usepackage{booktabs}
\usepackage{ifthen}
\usepackage{pifont}
\usepackage{gensymb}
\usepackage{nicefrac}
\usepackage{blindtext}
\usepackage{xcolor}
\usepackage{subcaption}
\usepackage{multirow}
\usepackage{cite}
\usepackage[nolist,nohyperlinks]{acronym}

\usepackage{xkeyval}

\interspeechcameraready

\title{Spatial LibriSpeech: An Augmented Dataset for Spatial Audio Learning}
\name{Miguel~Sarabia, Elena~Menyaylenko, Alessandro~Toso, Skyler~Seto, Zakaria~Aldeneh, Shadi~Pirhosseinloo, Luca Zappella, Barry-John~Theobald, Nicholas~Apostoloff, Jonathan~Sheaffer}
\address{Apple}
\email{miguelsdc@apple.com}

\newcommand{\sqm}{m\(^2\)}
\newcommand{\st}{\(^\mathrm{st}\)}

\newcommand{\rd}{\(^\mathrm{rd}\)}
\renewcommand{\th}{\(^\mathrm{th}\)}
\newcommand{\mdeg}{\(\degree\) }

\begin{acronym}
\acrodef{DRR}{direct-to-reverberant ratio}
\acrodef{RIR}{room impulse response}
\acrodef{MRP}{mouth reference point}
\acrodef{FID}{Fréchet inception distance}
\acrodef{SNR}{signal-to-noise ratio}
\acrodef{MLP}{multi-layer perceptron}
\end{acronym}

\begin{document}
\maketitle

\begin{abstract}
We present Spatial LibriSpeech, a spatial audio dataset with over 650 hours of 19-channel audio, first-order ambisonics, and optional distractor noise. Spatial LibriSpeech is designed for machine learning model training, and it includes labels for source position, speaking direction, room acoustics and geometry. Spatial LibriSpeech is generated by augmenting LibriSpeech samples with 200k+ simulated acoustic conditions across 8k+ synthetic rooms. To demonstrate the utility of our dataset, we train models on four spatial audio tasks, resulting in a median absolute error of 6.60\mdeg on 3D source localization, 0.43m on distance, 90.66ms on T30, and 2.74dB on \acl{DRR} estimation. We show that the same models generalize well to widely-used evaluation datasets, e.g., obtaining a median absolute error of 12.43\mdeg on 3D source localization on TUT Sound Events 2018, and 157.32ms on T30 estimation on ACE Challenge.

\end{abstract}

\acresetall 

\section{Introduction}
\label{sec:intro}

Humans can infer a great amount of information from their acoustic environment, including determining the spatial location of sound sources, estimating the size of a room, and estimating the room reverberance \cite{ohuchi2006, hameed2004}. To develop machine learning algorithms with the same level of acoustic spatial awareness, we require a large and diverse dataset with annotations for each task. To the best of our knowledge, no publicly-available dataset has the scale, and diversity needed to train general models for multiple spatial audio tasks. This lack of large-scale datasets limits the applicability of modern machine learning techniques to spatial audio.

In this paper, we introduce Spatial LibriSpeech\footnote{See  \href{https://github.com/apple/ml-spatial-librispeech}{\texttt{github.com/apple/ml-spatial-librispeech}}.}, a spatially augmented version of LibriSpeech~\cite{LibriSpeech} with optional noise from the Microsoft Deep Noise Suppression Challenge 2021~\cite{MS_DNS}. Spatial LibriSpeech augments the LibriSpeech and Microsoft Deep Noise Suppression samples by simulating how they would be perceived by a microphone array in various synthetic rooms. Spatial LibriSpeech contains over 650 hours of spatial audio with labels for source position, speaking direction, room acoustics, and room geometry (refer to Section \ref{sec:background} for a review of spatial audio tasks for which these labels are useful). Our goal is for Spatial LibriSpeech to be the main training dataset for spatial audio applications.

To create Spatial LibriSpeech, we first generated 8,952 synthetic rooms, which were used to obtain \acp{RIR} that were convolved and scaled with LibriSpeech samples. The \acp{RIR} are modeled on the Zylia microphone array, a 19-channel spherical microphone array. This array facilitates a means for extracting a third-order ambisonics representation up to a frequency of 3080Hz~\cite{Pinardi2021}. For completeness, we also provide synthesized, full-bandwidth first-order ambisonics~\cite{ambisonics}, which are aliasing free. Either of these formats may be used to simulate a wide variety of arrays~\cite{gamper16}. We describe the dataset generation process in more detail in Section~\ref{ssec:dataset_generation}. A key advantage of convolving existing speech samples with synthetic \acp{RIR} for spatial audio dataset generation is that we can ensure our dataset spans a variety of acoustic room properties and room sizes. For instance, full-band T30 values range from 145ms to 2846ms, and room floor area ranges from 13.3\sqm~to 277.4\sqm. Section~\ref{ssec:dataset_statistics} describes diversity statistics of Spatial LibriSpeech.

While Spatial LibriSpeech has the potential to be used to train models for multiple tasks, in this paper we focus on some of the most fundamental spatial audio detection tasks: i) 3D source localization, ii) source distance, iii) third-octave narrow-band \acp{DRR}, and iv) third-octave narrow-band T30s. All models share the same architecture and training regime (presented in Section~\ref{sec:benchmarks}). When training spatial audio models, there is a choice of using the microphone array signals as inputs or converting the microphone array signals to ambisonics to obtain device-agnostic models. In our work, models consume first-order ambisonics.

Our models achieve a median absolute error of 6.60\mdeg in 3D source localization, 0.43m in distance estimation, 2.74dB in \ac{DRR} estimation, and 90.66ms in T30 estimation. Since our training dataset is composed of simulated acoustics, we verify the transferability of our models by testing them on two evaluation datasets: TUT Sound Events 2018~\cite{TUTSoundEvents}, and ACE Challenge~\cite{ACE} (Section~\ref{ssec:transferability}). We find that fine-tuning is beneficial for ACE Challenge but unnecessary for TUT Sound Events 2018, and we plot the responses of our models on both evaluation datasets showing that ACE responses are at the tail of the Spatial LibriSpeech responses (see Section~\ref{ssec:embeddings}). Lastly, we show that, for single-task models, using a version of Spatial LibriSpeech made by uniformly sampling 10\% of the dataset yields the same performance as the full dataset (cf. Section~\ref{ssec:lite_experiments}). 


\section{Background}
\label{sec:background}
\begin{table*}[t]
\scriptsize
\centering

\caption{Comparison of existing datasets. \textsc{Rooms} refers to the number of unique acoustic environments in the dataset, \textsc{Room Config.} to the number of physical configurations per environment (e.g. different source receiver positions), \textsc{Channels} to the availability of spatial information from a microphone array or other encoding: N - N channels mic array, B - Binaural, X\(^\mathit{th}\)OA - X\(^\mathit{th}\) order ambisonics, \textsc{Data-types}: sIR - Simulated IR, rIR - Recorded IR, sA - Simulated Audio, rA - Recorded Audio. \textsc{Labels}: P - Position, SD - Speaking direction, R - Room acoustics, G - Room geometry, O - Other.}

\label{table:spatial_audio_datasets}

\begin{tabular}{
>{\raggedleft\arraybackslash}m{0.40\columnwidth}%
>{\centering\arraybackslash}m{0.25\columnwidth}%
>{\centering\arraybackslash}m{0.25\columnwidth}%
>{\centering\arraybackslash}m{0.25\columnwidth}%
>{\centering\arraybackslash}m{0.25\columnwidth}%
>{\centering\arraybackslash}m{0.25\columnwidth}%
}

\toprule
\textsc{Dataset} & \textsc{Rooms} & \textsc{Room Config.} & \textsc{Channels} & \textsc{Data-types} & \textsc{Labels} \\ 
\midrule

dEchorate~\cite{dEchorate} & 11 & 180 & 5 & sIR, rIR & P \\ 

Arni~\cite{Arni} & 1 & 21 & 3\rd OA, 4\th OA & rIR & P, R \\ 

GIR~\cite{GIR}& 1 & 2,951 & 1 & rIR & P, O \\ 

EasyCom~\cite{EasyCom} & 1 & 50 & 4, B & rA & P, R, G, O \\ 

CoupledRooms~\cite{CoupledRooms} & 2 & 101 & 4\th OA & rIR & P, R, G \\ 

DCASE2021 task 3~\cite{DCASE} & 13 & 1,184--6,480 & 4, 1\st OA & rIR, rA & P \\ 

BUT ReverbDB~\cite{ReverbDB} & 8 & 155 & 1 & rIR & P, R, G \\ 

SBSBRIR~\cite{SBSBRIR} & 1 & 180 x $2\degree$ & B & rIR & P \\ 

BIRD~\cite{BIRD} & 12,500 & 8 & 2 & sIR & P, R \\ 

Kemar BRIRs~\cite{KemarBRIRs} & 1 & 50 & B & rIR & P, G \\ 

Motus~\cite{Motus} & 1 & 3,320 & 32, 4\th OA & rIR & P, R, G \\ 

Aachen IR database~\cite{Aachen}& 4 & 17 & B & rIR & P, R, G, O \\ 

ACE Challenge~\cite{ACE} & 7 & 10 & 2 - 32 & rIR & P, R \\ 

DIHRA~\cite{DIRHA} & 2 & 62 & 3 & rA, sA & P, G \\ 

Voice-Home~\cite{VoiceHome} & 12 & 24 & 8 & rIR, rA & P, G \\ 

Sweet-Home~\cite{SweetHome} & 4 & 7 & 1 & rA & - \\ 

Microsoft DNS 2001~\cite{MS_DNS} & 14,576 & 1 & 1 & sIR, rIR & R \\ 

TUT Sound Events 2018~\cite{TUTSoundEvents} & 5 & 487-4,366 & 8, 1\st OA & sA, rA & P, G \\
\midrule 
\textbf{Spatial LibriSpeech} & 8,952 & 20 & 19, 1\st OA & sA & P, SD, R, G \\ 
\bottomrule
\end{tabular} 
\vspace*{-0.3cm}
\end{table*}

We start by reviewing \textit{spatial audio} tasks to highlight the large number of potential use cases for Spatial LibriSpeech. We divide audio tasks into three categories: \textit{source parameter estimation}, \textit{environment parameter estimation}, and \textit{spatial processing}. Source parameter estimation tasks are concerned with understanding audio sources. Examples include estimation of: source localization~\cite{Grumiaux2022}, source distance~\cite{Gburrek2020}, and speaking direction~\cite{Ahuja2020}. Environment parameter estimation tasks involve understanding the environment where the audio is produced. Example tasks include estimation of \ac{DRR}~\cite{Bryan2020}, material absorption and scattering~\cite{Pelzer2013}, and room volume~\cite{Srivastava2021}. Finally, spatial processing tasks involve the transformation of audio signals with information extracted from the acoustic environment. Example tasks include de-reverberation~\cite{Kinoshita2017}, beamforming~\cite{Heymann2016}, and audio source separation~\cite{Tzinis2020}.

Table~\ref{table:spatial_audio_datasets} contains a summary of the main differences between 18 published spatial audio datasets in terms of the number of environments, number of physical configurations per environment, recorded/simulated channels, and data-types and labels included in the dataset. We found only two datasets that include over 50k unique configurations across all environments: BIRD~\cite{BIRD} and DCASE2021 Task 3~\cite{DCASE}. However, BIRD features only two microphone channels, and DCASE2021 Task 3 includes only position labels. Spatial LibriSpeech is the only dataset to feature over 200k unique configurations, labels for a large number of acoustic tasks, and both mic-domain audio and ambisonics. This lack of diversity and labels limits the applicability of existing datasets to modern machine learning techniques~\cite{Gong2019} such as multi-task learning or contrastive representation learning~\cite{data2vec}.

\section{The Spatial LibriSpeech Dataset}
\label{sec:spatial_librispeech}
In this section, we describe the generation of Spatial LibriSpeech and its defining characteristics.

\subsection{Generation Methodology}
\label{ssec:dataset_generation}

Our pipeline to generate Spatial LibriSpeech consists of three steps: \textit{parametric room generation}, \textit{room impulse response simulations}, and \textit{mixing}.

\textit{Parametric room generation.} We start by defining a realistic set of conditions for commonly encountered living spaces following~\cite{Diaz2005}. These consist of a predefined distribution of reverberation times and room shapes and sizes. The materials associated with the room surfaces are chosen from a database of typical construction materials, for which the absorption and scattering coefficients as a function of frequency are available. 

\textit{Room impulse response simulations.} We use a geometrical acoustic solver, which includes a high fidelity model of the room that accounts for the directivity of the microphones (including diffraction effects from the array body). The room is populated with several acoustic objects including the recording device placed at a randomized but bounded position, several sources surrounding the device with a randomized looking direction and directivity function. When sources are intended to represent speech, their directivity is computed using a boundary element code using a set of artificially generated  models. The direct path from each source to the microphone array, the early reflections, and the late reverberations are used to assemble each impulse response. 

\textit{Mixing.} Once we have computed \acp{RIR} for each configuration of the simulated rooms, we mix them with LibriSpeech samples by randomly assigning them to sources in every simulated room. We then remove leading and trailing silence from each sample\footnote{Silence is defined as any leading or trailing sound before a 100ms segment at more than 1\% volume.}, convolve the sample with the selected \ac{RIR} and scale the output to a random active speech level (ASL) between 85dB-ASL and 100dB-ASL at mouth reference point\footnote{Mouth reference point is the point on the reference axis 25 mm in front of the lip plane ITU-T P.581~\cite{ITU-T-P.581}}. Finally, we save the resulting audio samples and store the relevant labels from both the room simulation and LibriSpeech. For the distractor noise we follow the same process, except we assign a random source, different to the main signal source, and set the signal strength to a random \ac{SNR} between 10dB and 40dB.

\subsection{Dataset Characteristics}
\label{ssec:dataset_statistics}

\begin{table}[t]
\scriptsize
\centering
\caption{Main characteristics of Spatial LibriSpeech.}
\hspace*{-0.3cm}
\begin{tabular}{
>{\raggedleft\arraybackslash}m{0.27\columnwidth}%
>{\centering\arraybackslash}m{0.30\columnwidth}%
>{\centering\arraybackslash}m{0.30\columnwidth}%
}
\toprule

& 
\textsc{Train} &
\textsc{Test} \\

\midrule

Total duration & 573h 13m 12s & 85h 29m 20s \\ 

Speech samples & 171,951 & 49,505\\


Simulated rooms & 8,952 & 4,970 \\

Source azimuth & %
    [-180.0\mdeg, +180.0\(\degree\)] & %
    [-180.0\mdeg, +180.0\(\degree\)]\\

Source elevation & %
    [-48.1\mdeg, +48.7\(\degree\)] &
    [-49.5\mdeg, +42.4\(\degree\)] \\

Source distance & %
    [0.5m, 4.0m] &%
    [0.9m, 4.0m] \\

Speaking azimuth & %
    [-180.0\mdeg, +180.0\(\degree\)] & %
    [-180.0\mdeg, +180.0\(\degree\)]\\

Speaking elevation & %
    [-89.3\mdeg +87.7\(\degree\)] & 
    [-74.5\mdeg, +68.2\(\degree\)]\\

Voice directivities & 16 variations & 1 variation \\

Room floor area & %
    [13.3\sqm, 277.4\sqm] &
    [14.3\sqm, 277.4\sqm] \\ 

Full-band T30 & %
    [144.5ms, 2846.0ms] &
    [157.8ms, 1267.7ms] \\
 
\bottomrule
\end{tabular}
\label{table:spatial_librispeech}
\vspace*{-0.3cm}
\end{table}

The main characteristics of Spatial LibriSpeech are summarized in Table~\ref{table:spatial_librispeech}. The dataset consists of two splits: \textsc{train} which is derived from LibriSpeech's \texttt{train-clean-100} and \texttt{train-clean-360} subset, and \textsc{test} which is derived from \texttt{test-clean}. The audio is sampled at 16kHz.

Spatial LibriSpeech includes the following labels: source localization (azimuth and elevation), speaking direction, room volume, surface and floor area, voice directivity identifier, narrow-band C50, \ac{DRR}, EDT, T20, and T30~\cite{ISO-3382-1}; and all original metadata from LibriSpeech, including ids for speaker, books, and chapters, and the reference transcriptions.

We verify that our models do not overfit to the virtual rooms in the training set by checking that the performance of models trained with Spatial LibriSpeech transfer to two existing evaluation datasets (see Section~\ref{ssec:transferability}).

\section{Training with Spatial LibriSpeech}
\label{sec:benchmarks}
To demonstrate the utility of Spatial LibriSpeech in real-world situations, we trained several neural networks for: i) 3D source localization, ii) source distance, iii) \ac{DRR}, and iv) T30. We chose these tasks since they are representative of the source and environment parameter estimation spatial audio tasks and because public third party evaluation datasets were available to analyse transferability.

For 3D source localization, we report the median absolute error of the 3D angle (cf.~\cite{pilot2021}):
\begin{equation}
\footnotesize
    \alpha = \cos^{-1}\left(\sin(\phi)\sin(\hat{\phi})+\cos(\phi)\cos(\hat{\phi} )\cos(\theta - \hat{\theta})\right),
\label{eq:3d_angle}
\end{equation}
where \(\phi\)  and \(\theta\) are the ground-truth azimuth and elevation respectively, and \(\hat{\phi}\) and \(\hat{\theta}\) are the predicted azimuth and elevation. For distance regression, we report the median absolute error in meters. As \ac{DRR} and T30 are 20-dimensional vectors representing the third-octave frequencies between 100Hz and 8kHz, we report both the median absolute error as well as the Pearson correlation across frequencies.

Our objective is to show that models trained with Spatial LibriSpeech transfer to two established evaluation datasets:  ACE Challenge~\cite{ACE} and TUT Sound Events 2018~\cite{TUTSoundEvents}. All models share the same architecture. We first transform and normalize segments\footnote{The duration of the segments varies by task, for 3D source localization and distance we use 0.5s, for T30 and \ac{DRR} we use 4.0s.} of spatial audio encoded as 4-channel first-order ambisonics into active and reactive components following~\cite{Jacobsen1990}. The active and reactive components are fed through two independent branches of four 3D-convolutional layers each (of 2, 4, 8 and 16 channels respectively), with max-pool, batch-norm and exponential linear units between the convolutional layers. Next, the output of both branches is flattened and concatenated, and fed to a 3-layer \ac{MLP}. The number of outputs of the last layer depends on the task: two for source localization (one for azimuth, and another for elevation), one for distance regression, 20 for \ac{DRR}, and 20 for T30. Our model has 101,933 parameters (for 3D source localization and distance) and 747,148 parameters (for T30 and DRR). Models are trained for 20 epochs for 3D source localization and distance, and 50 epochs for T30 and DRR\footnote{This represents 80.8M gradient updates for 3D source localization and distance, and 21.2M gradient updates for T30 and DRR estimation.}. We use the Adam optimizer, a weight decay of 0.01, dropout of 50\% on the output of the convolutional block, Kaiming uniform initialization, and a learning rate of 10\textsuperscript{-5}. Our models completed training in a median 8h15min for 3D Source Localization and Distance Estimation, and 20h for T30 and \ac{DRR} on a 80-core Intel Xeon CPU with 8 NVIDIA Tesla V100s GPUs, and 375GiB of RAM.

\subsection{Performance and transferability}
\label{ssec:transferability}

\newcommand{\bestmae}{\(\tilde{\epsilon_\star}\)}
\newcommand{\bestiqr}{\(\Delta_\star\)}
\newcommand{\modelsiqr}{\(\Delta_\mathrm{m}\)}
\newcommand{\bestcorr}{\(\tilde{\rho_\star}\)}

\makeatletter 
\define@key{transferresults}{bestmae}{\gdef\transferresults@bestmae{#1}}
\define@key{transferresults}{bestiqr}{\gdef\transferresults@bestiqr{#1}}
\define@key{transferresults}{modelsiqr}{\gdef\transferresults@modelsiqr{#1}}
\define@key{transferresults}{bestcorr}{\gdef\transferresults@bestcorr{#1}}
\define@key{transferresults}{unit}{\gdef\transferresults@unit{#1}}
\define@key{transferresults}{figure}{\gdef\transferresults@figure{#1}}

\newcommand{\vecresults}[1]{%
    \setkeys{transferresults}{#1} %
    {%
    \bestmae:~{\transferresults@bestmae}{\transferresults@unit} %
    \bestiqr:~{\transferresults@bestiqr}{\transferresults@unit} %
    \bestcorr:~{\transferresults@bestcorr} %
    \modelsiqr:~{\transferresults@modelsiqr}{\transferresults@unit} %
    \includegraphics[width=\linewidth]{\transferresults@figure}%
    }%
}

\newcommand{\scalarresults}[1]{%
    \setkeys{transferresults}{#1} %
    {%
    \bestmae:~{\transferresults@bestmae}{\transferresults@unit} %
    \bestiqr:~{\transferresults@bestiqr}{\transferresults@unit} %
    \modelsiqr:~{\transferresults@modelsiqr}{\transferresults@unit} %
    \includegraphics[width=\linewidth]{\transferresults@figure}%
    }%
}

\makeatother 

\begin{table}[t!]
\scriptsize
\caption{Performance of models trained with the Spatial LibriSpeech on 3D Sound Localization, Distance, T30 and DRR regression. For each task, 5 models are trained with the train set of Spatial LibriSpeech and evaluated against the test set of Spatial LibriSpeech, TUT Sound Events 2018~\cite{TUTSoundEvents}, and ACE Challenge~\cite{ACE}. We report the median absolute error (\bestmae) and IQR across predictions (\bestiqr) for the best performing model on the test set of Spatial LibriSpeech. We also report the IQR across the median absolute error of each model (\modelsiqr). For DRR and T30, \bestmae, \bestiqr, and \modelsiqr~are taken over all frequency bins. Additionally, for DRR and T30, we report the median Pearson correlation of the best performing model across frequency bin for each prediction (\bestcorr). Purple violin plots show the density of the absolute error, while pink violin plots show the density of the Pearson correlations, both for the best performing models.}
\label{table:transfer_results}
\centering
\hspace*{-0.2cm}
\begin{tabular}{%
>{\raggedleft\arraybackslash}m{0.18\columnwidth}%
>{\centering\arraybackslash}m{0.36\columnwidth}%
>{\centering\arraybackslash}m{0.36\columnwidth}%
}
\toprule
& \textsc{3D Source Local.} & \textsc{Distance Estimation} \\

\midrule

\textsc{Spatial LibriSpeech (Test)} &%
    \scalarresults{ 
        bestmae=6.60,
        bestiqr=7.75,
        modelsiqr=0.52,
        unit=\mdeg,
        figure=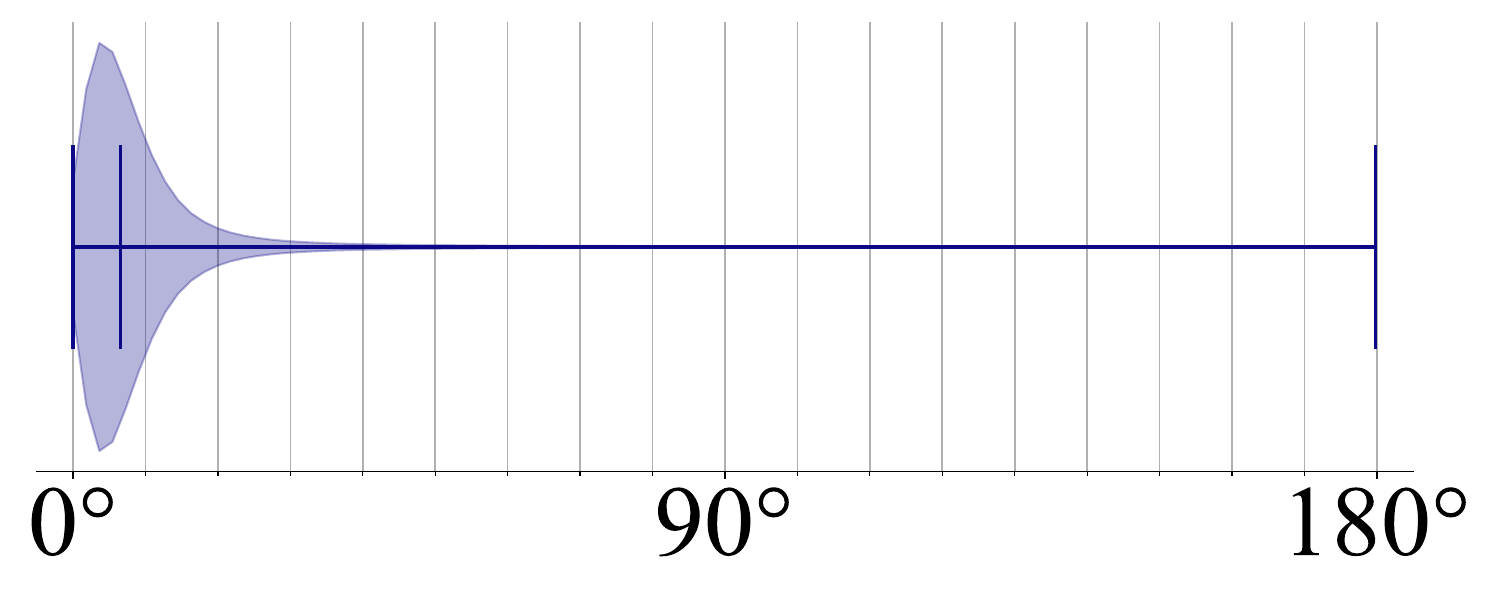
    } & 
    \scalarresults{ 
        bestmae=0.43,
        bestiqr=0.56,
        modelsiqr=0.01,
        unit=m,
        figure=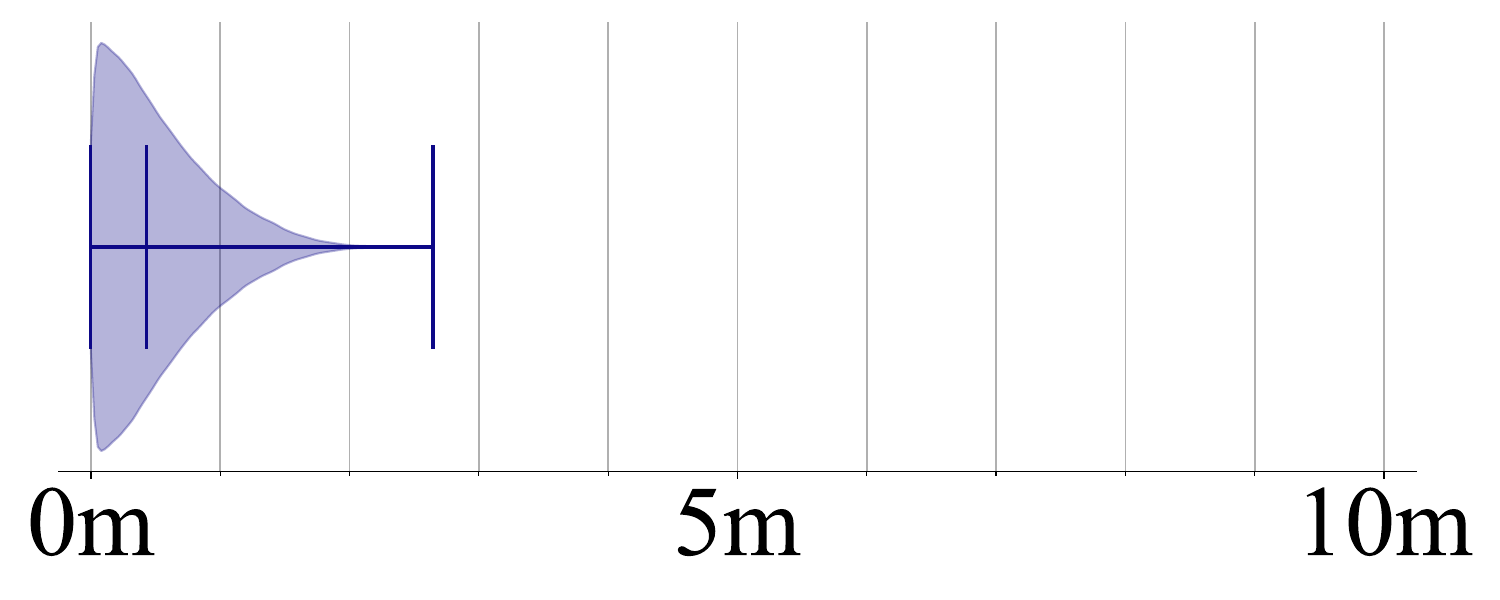
    } \\

\cmidrule{2-3}

\textsc{TUT Sounds (ANSIM)} &%
    \scalarresults{ 
        bestmae=8.19,
        bestiqr=9.49,
        modelsiqr=1.82,
        unit=\mdeg,
        figure=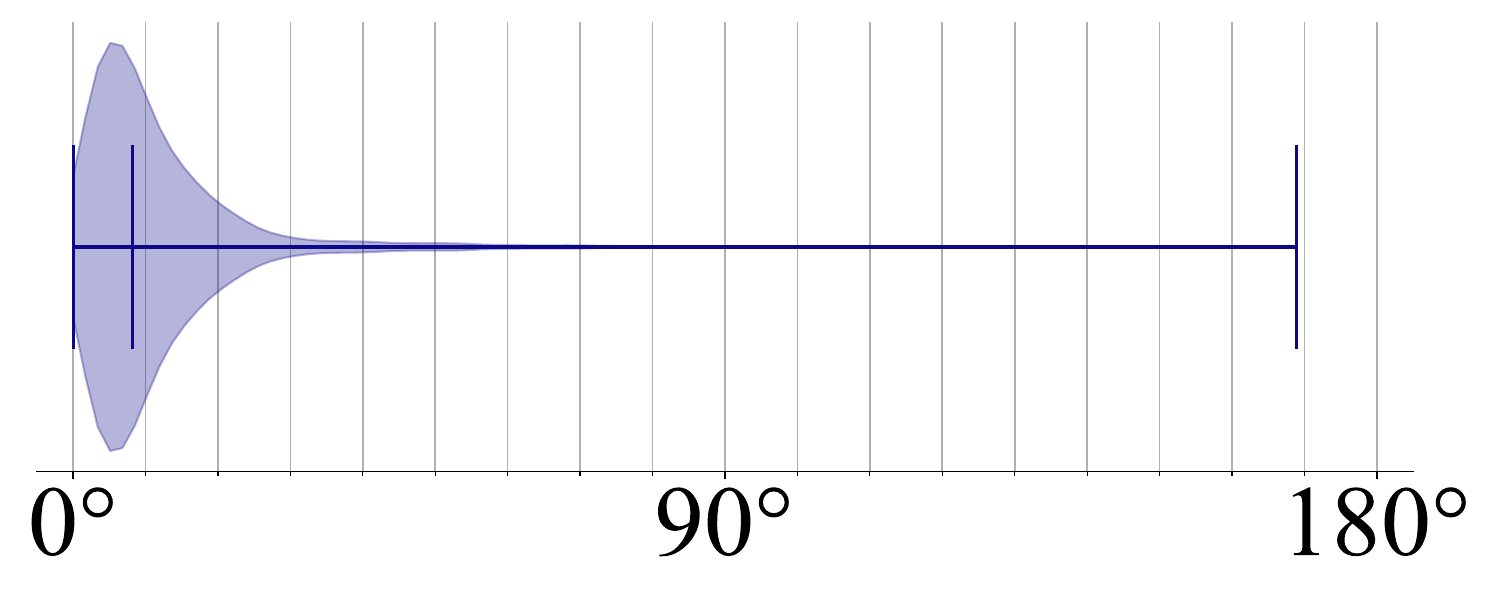
    } & 
    \scalarresults{ 
        bestmae=4.76,
        bestiqr=4.39,
        modelsiqr=0.04,
        unit=m,
        figure=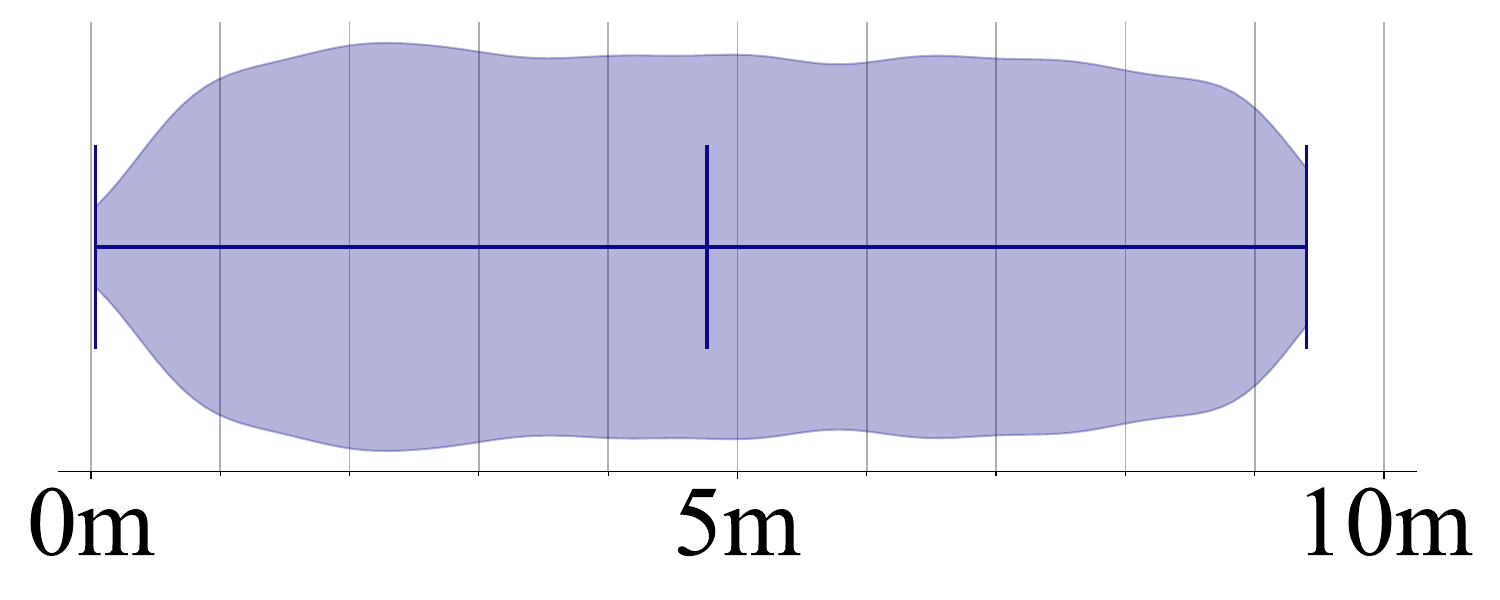
    } \\

\cmidrule{2-3}

\textsc{TUT Sounds (RESIM)} & %
    \scalarresults{ 
        bestmae=5.80,
        bestiqr=5.57,
        modelsiqr=0.51,
        unit=\mdeg,
        figure=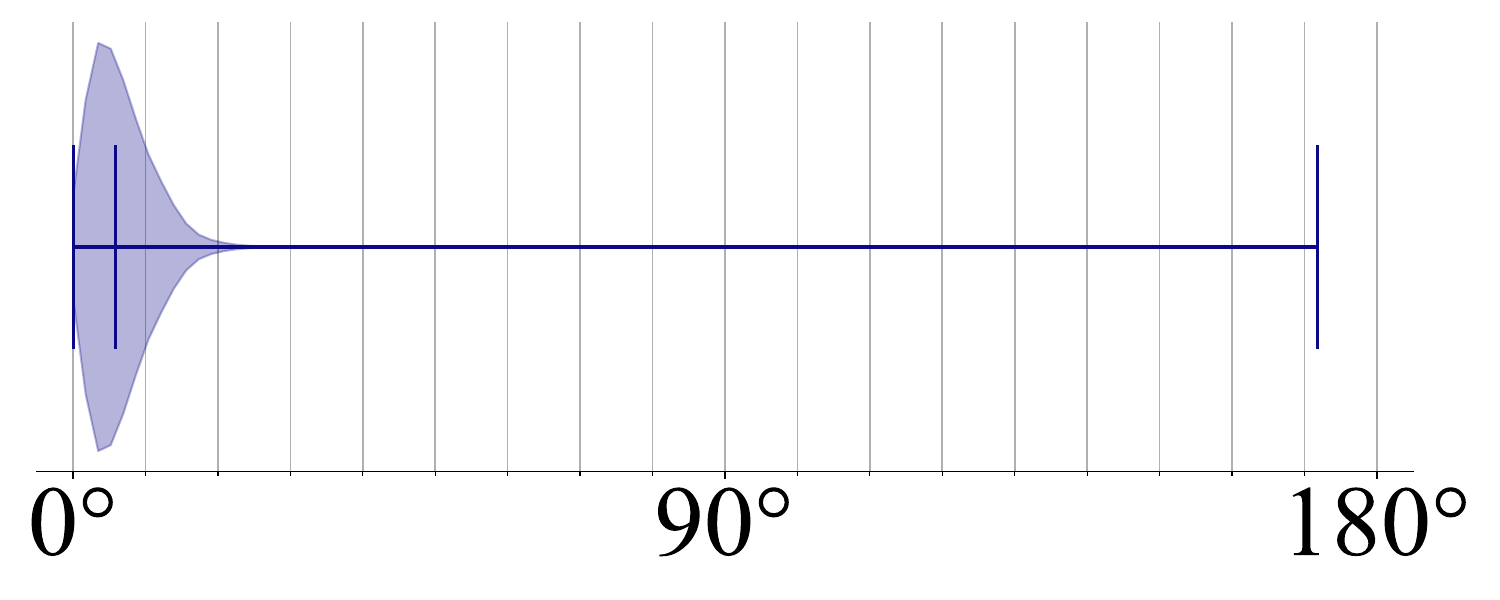
    } & 
    \scalarresults{ 
        bestmae=0.25,
        bestiqr=0.30,
        modelsiqr=0.02,
        unit=m,
        figure=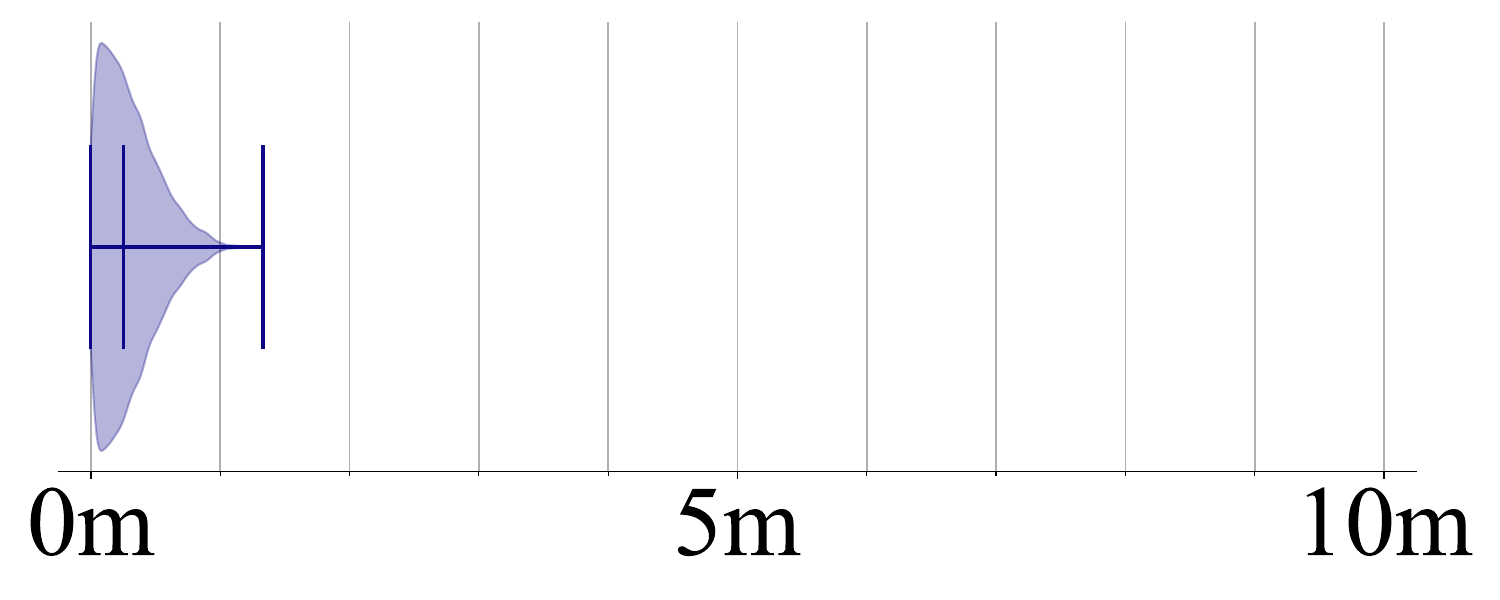
    } \\

\cmidrule{2-3}

\textsc{TUT Sounds (REAL)} &  %
    \scalarresults{ 
        bestmae=12.43,
        bestiqr=15.96,
        modelsiqr=1.01,
        unit=\mdeg,
        figure=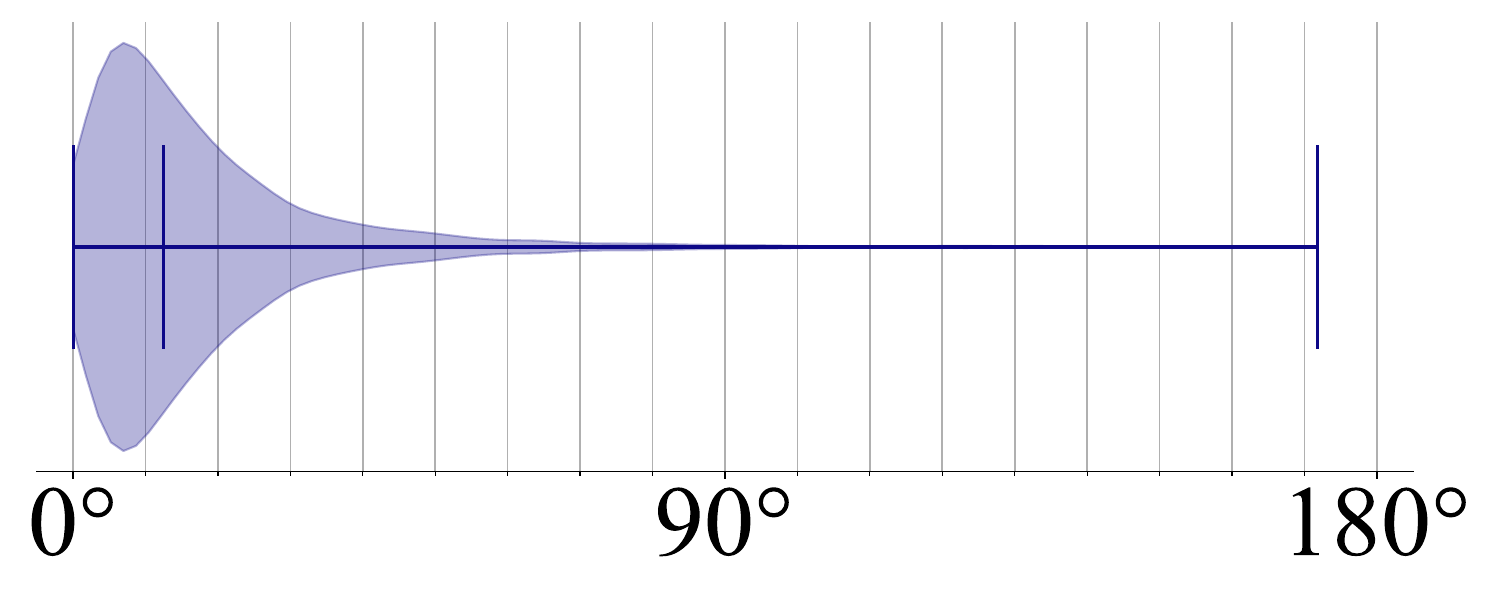
    } & 
    \scalarresults{ 
        bestmae=0.60,
        bestiqr=0.85,
        modelsiqr=0.03,
        unit=m,
        figure=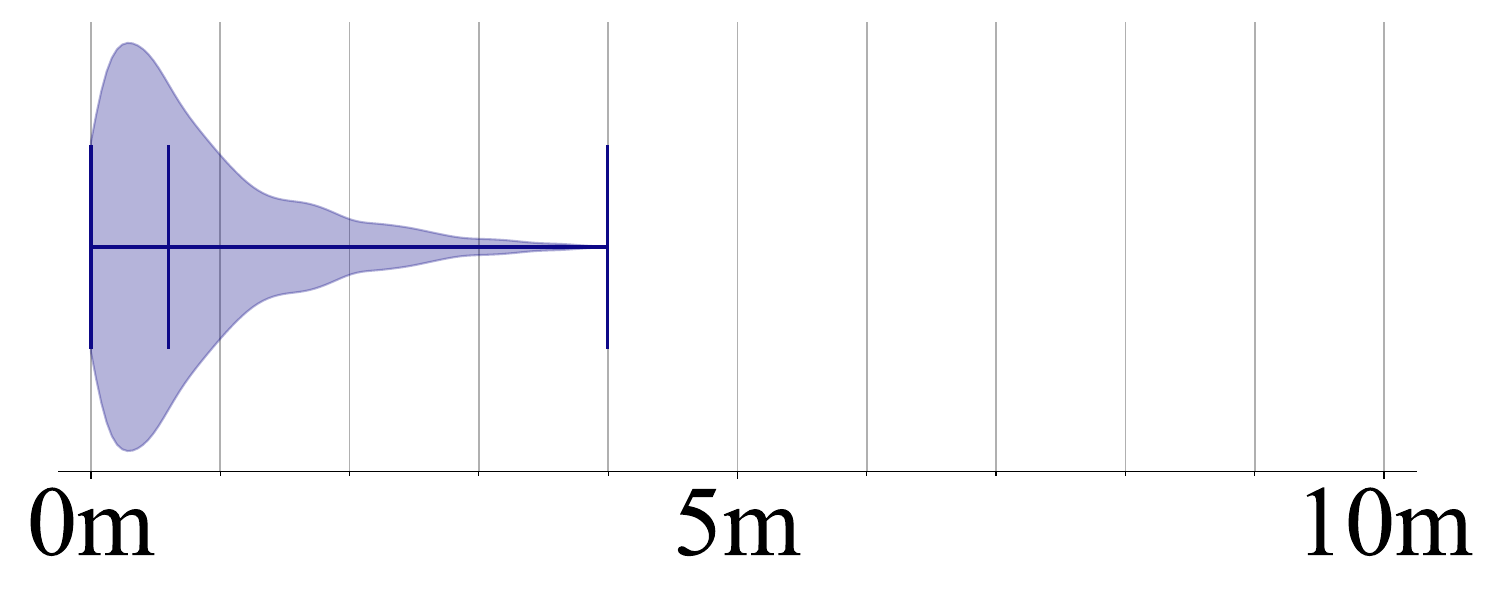
    } \\

\midrule

& \textsc{DRR} & \textsc{T30} \\

\midrule

\textsc{Spatial LibriSpeech (Test)} &%
    \vecresults{ 
        bestmae=2.74,
        bestiqr=3.49,
        modelsiqr=0.35,
        bestcorr=0.98,
        unit=dB,
        figure=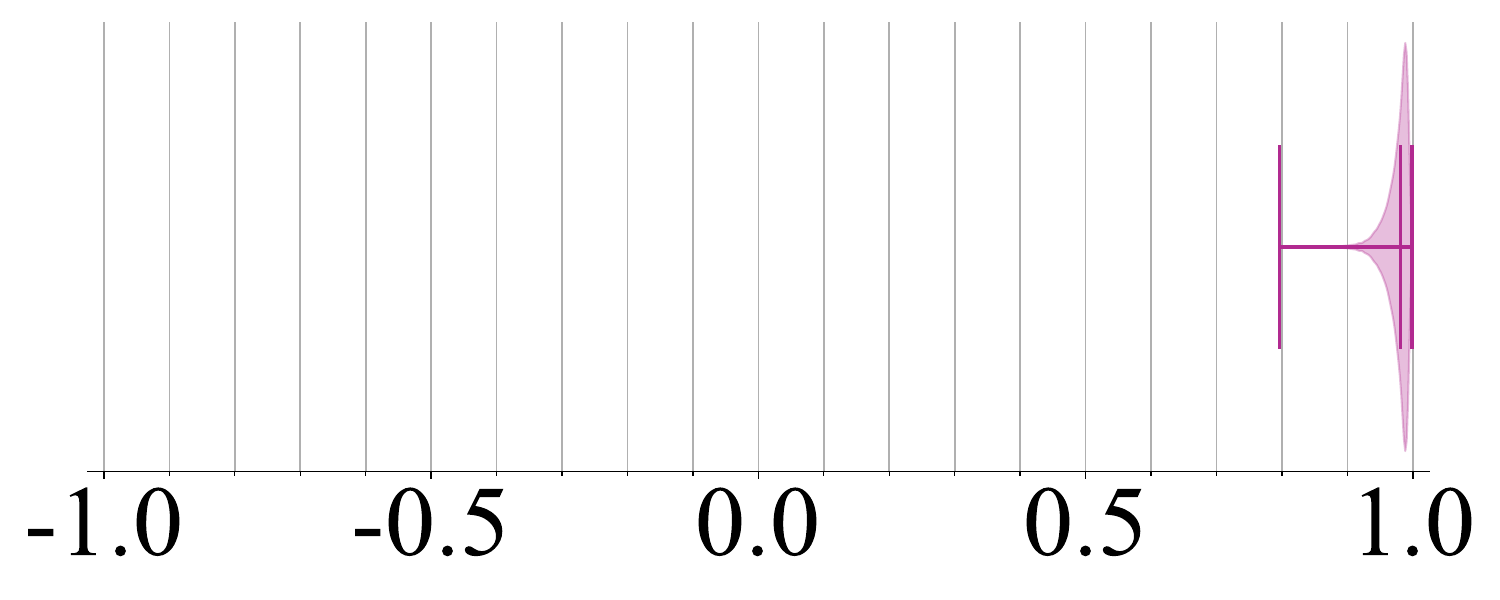
    } & 
    \vecresults{ 
        bestmae=90.66,
        bestiqr=116.83,
        modelsiqr=35.12,
        bestcorr=0.46,
        unit=ms,
        figure=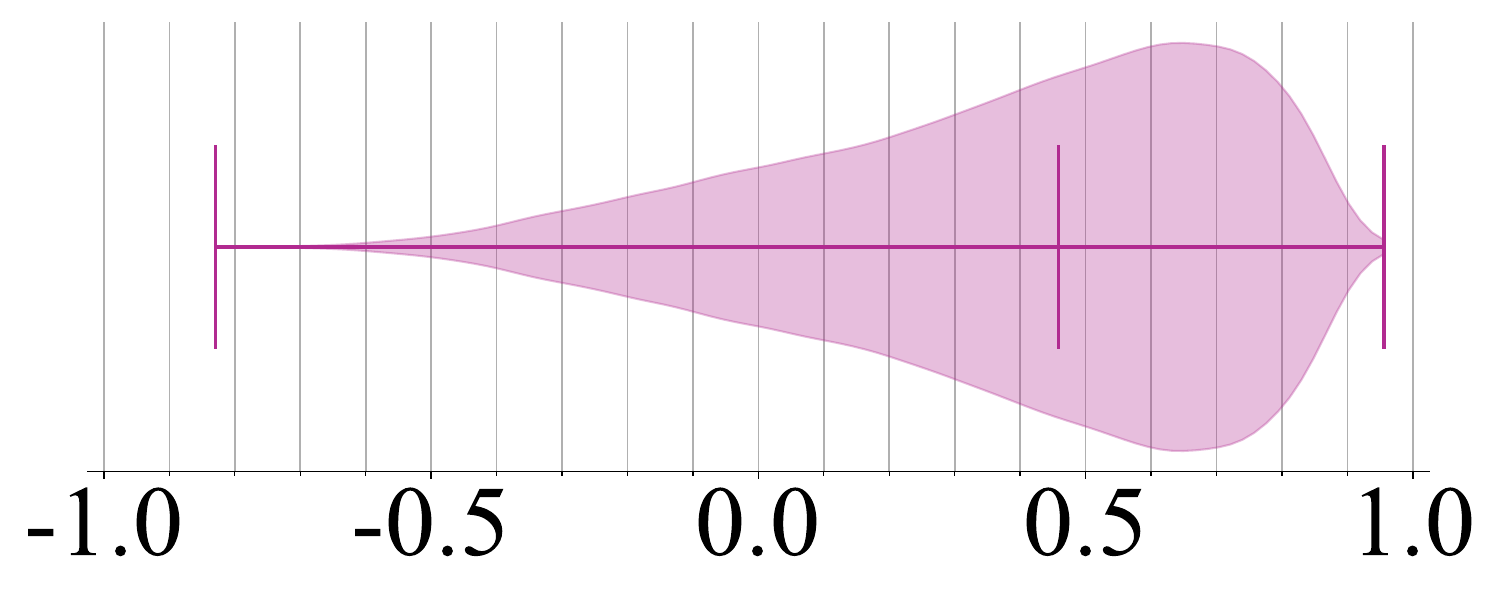
    } \\    

\cmidrule{2-3}

\textsc{ACE Challenge} &%
    \vecresults{ 
        bestmae=14.50,
        bestiqr=14.75,
        modelsiqr=1.40,
        bestcorr=0.74,
        unit=dB,
        figure=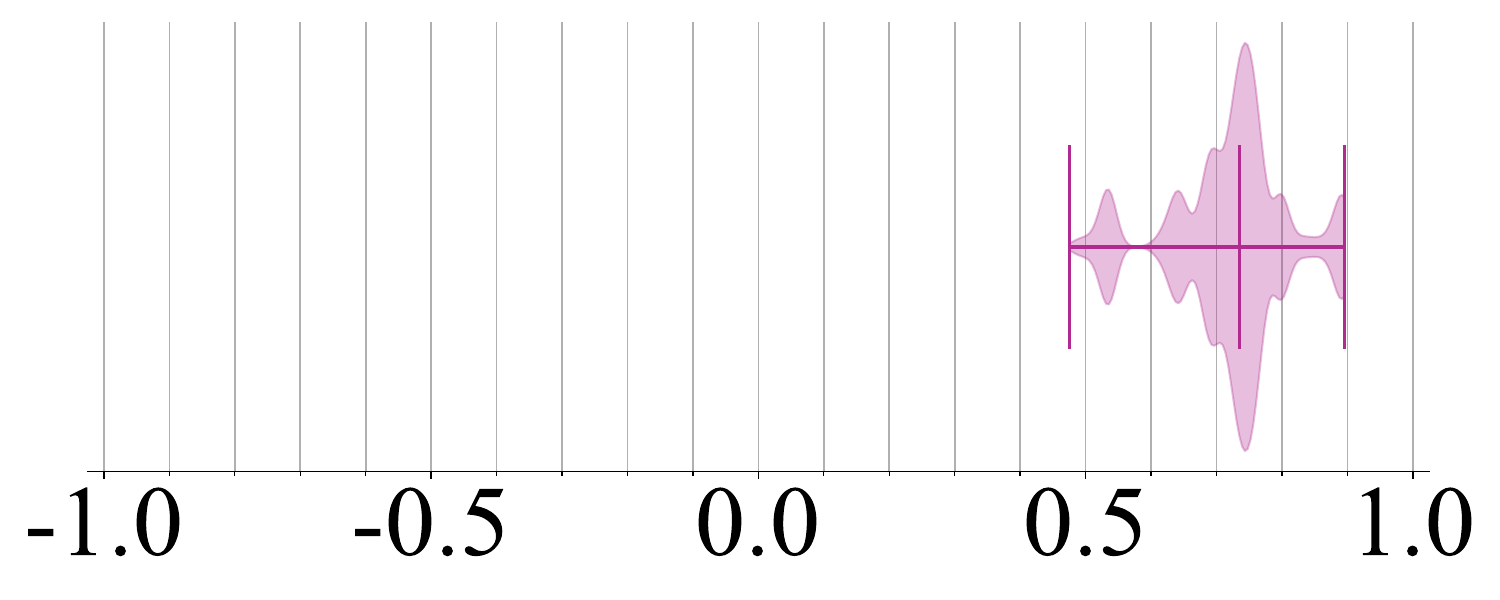
    } & 
    \vecresults{ 
        bestmae=157.32,
        bestiqr=230.13,
        modelsiqr=70.94,
        bestcorr=0.32,
        unit=ms,
        figure=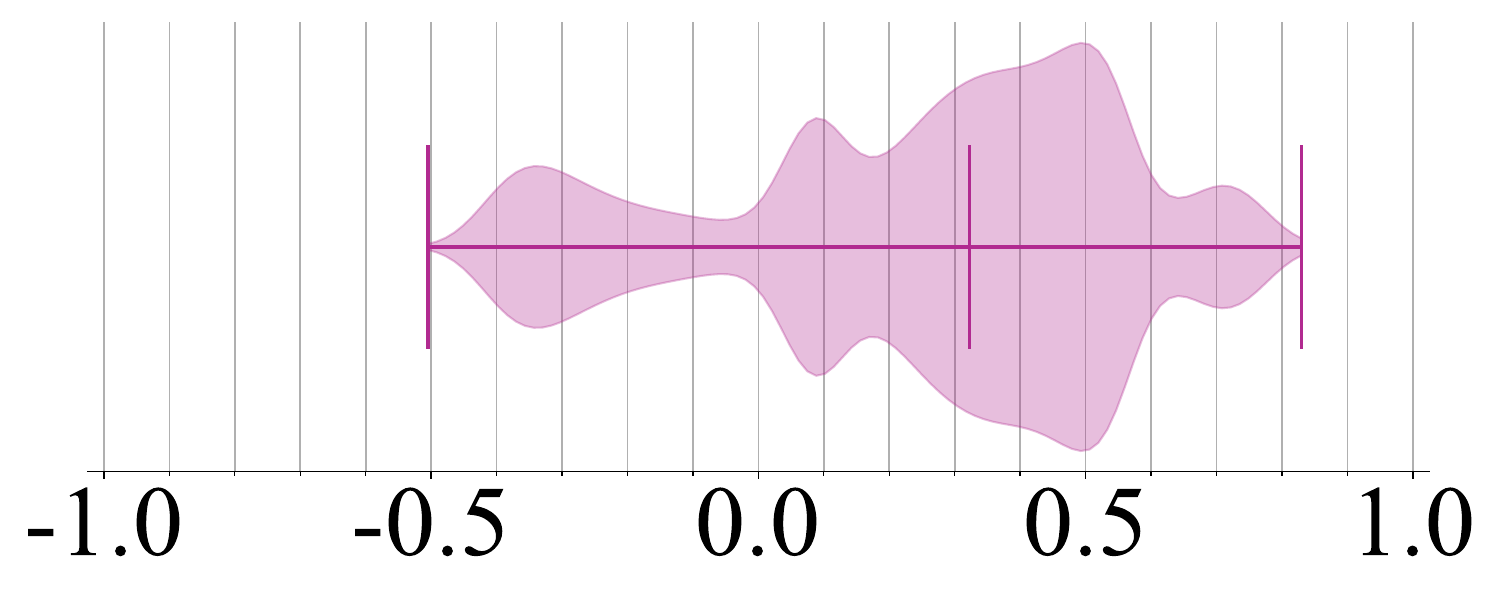
    } \\

\cmidrule{2-3}

\textsc{ACE Challenge (Finetuned)} &%
    \vecresults{ 
        bestmae=4.81,
        bestiqr=6.61,
        modelsiqr=3.79,
        bestcorr=0.40,
        unit=dB,
        figure=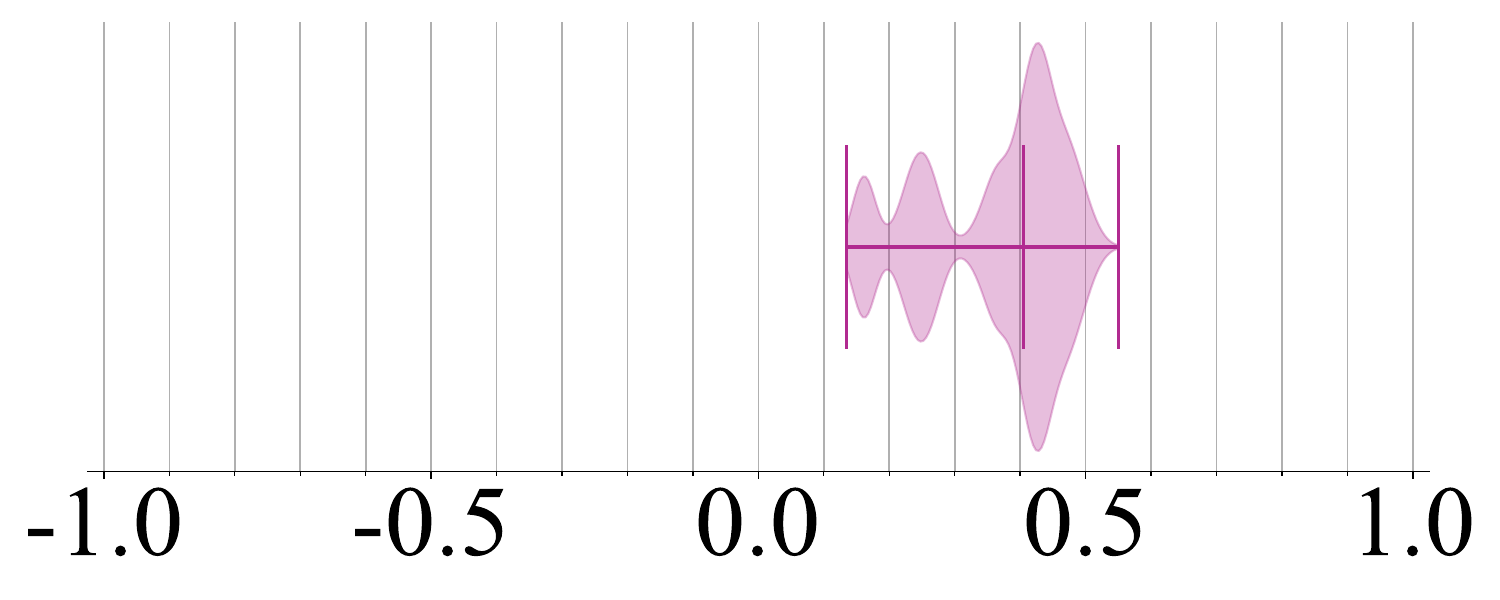
    } & 
    \vecresults{ 
        bestmae=105.57,
        bestiqr=187.28,
        modelsiqr=23.16,
        bestcorr=0.43,
        unit=ms,
        figure=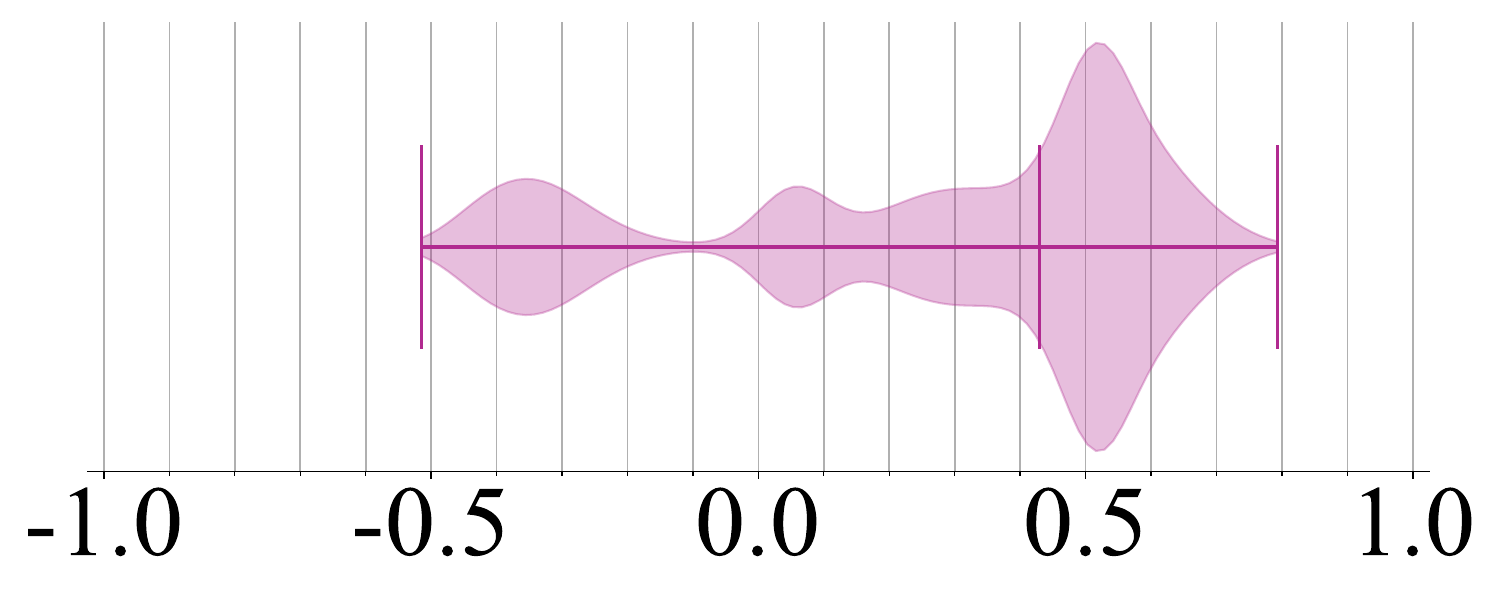
    } \\

\bottomrule
\end{tabular}
\vspace*{-0.5cm}
\end{table}

Table~\ref{table:transfer_results} shows the performance of our models on all four tasks when evaluated against the test set of Spatial LibriSpeech, TUT Sounds Events 2018~\cite{TUTSoundEvents}, and ACE Challenge Eval set\footnote{For ACE Challenge, we obtained first-order ambisonics from EM32 audio samples~\cite{rafaely2015}.}~\cite{ACE}. All models were trained using Spatial LibriSpeech only.

Our results show that the difference in performance between the test set of Spatial LibriSpeech and the performance on TUT Sound Events 2018 is small, +5.83\mdeg and +0.17m worse on REAL than Spatial LibriSpeech, though better on RESIM than on Spatial LibriSpeech (-0.90\mdeg and -0.18m). We find that our models do not exploit spurious correlations between the speech level and the distance labels, instead incorporating information from reverberations to make accurate predictions. This is evidenced by the median absolute error of ANSIM (an anechoic subset of TUT Sound Events 2018) being 4.76m.

Looking at the performance differences between Spatial LibriSpeech and ACE for DRR and T30 regression we find that the transfer penalty is higher (+11.76dB for DRR, and +66.66ms for T30). We also investigate fine-tuning the last three layers (the \ac{MLP}) with the ACE Challenge Dev set for 20 epochs, and find that the median absolute error improves significantly for both T30 and DRR (resulting in 2.07dB and 14.91ms gaps with the Spatial LibriSpeech test set).

For additional context into these results, Table~\ref{table:results_context} compares the performance of our models against the performance of a randomly initialized version of our model as well as the reported performance of recent benchmarks~\cite{pilot2021, Bryan2020}. In general, benchmarks obtain better results, as they use more complex architectures and train on data specifically modeled after the evaluation sets. We hypothesize that the performance of our model is close to the benchmarks due to the diversity of Spatial LibriSpeech, since we did not carry out an architecture or hyper-parameter search. Furthermore, note that none of these benchmarks published their training sets. In contrast, LibriSpeech was primarily designed for training spatial audio models.
\begin{table}[t]
\scriptsize
\caption{Performance contextualization for different tasks. \textsc{Random} refers to the performance of the model described in Section \ref{sec:benchmarks} without any training. \textsc{S.L.S} indicates the performance of our baseline (see Table~\ref{table:transfer_results}). \textsc{Ext.} refers to external benchmarks. The comparison with external benchmark may not be fair, as they were trained with different architectures, training regimes, and datasets closer to the evaluation dataset. \(^*\)Model was fine-tuned on ACE Dev.}
\label{table:results_context}
\centering
\hspace*{-0.4cm}
\begin{tabular}{%
>{\raggedright\arraybackslash}m{0.5cm}%
>{\centering\arraybackslash}m{1.2cm}%
>{\centering\arraybackslash}m{1cm}%
>{\centering\arraybackslash}m{1.2cm}%
>{\centering\arraybackslash}m{1.0cm}%
>{\centering\arraybackslash}m{1.5cm}%
}
\toprule
\textsc{Task} & \textsc{Eval.} & \textsc{Metric} & \textsc{Random} & \textsc{S.L.S.} & \textsc{Ext.} \\

\midrule

\multirow{3}{0.7cm}{3D Src Local.} & ANSIM\cite{TUTSoundEvents} & \multirow{3}{1.2cm}{\centering median abs. 3D angle error (eq.~\ref{eq:3d_angle})} & 88.49\mdeg & 8.19\mdeg  & \(\sim\)4.3\mdeg \cite{pilot2021}\\
 & RESIM\cite{TUTSoundEvents} & & 88.55\mdeg & 5.80\mdeg & \(\sim\)7.4\mdeg\cite{pilot2021}\\
 & REAL\cite{TUTSoundEvents} & & 88.53\mdeg & 12.43\mdeg & \(\sim\)4.3\mdeg\cite{pilot2021}\\
 \midrule
T30 & \multirow{2}{1cm}{ACE\cite{ACE}} & \multirow{2}{1.2cm}{\centering mean error} & -712.43ms & 156.44ms\(^\ast\) & 22.1ms \cite{Bryan2020}\\
DRR &  &  & 9.84dB & 4.04dB\(^\ast\) & 0.81dB \cite{Bryan2020}\\
\bottomrule

\end{tabular}
\vspace*{-0.2cm}
\end{table}

\subsection{Visualization of Dataset Representations}
\label{ssec:visualization}
\label{ssec:embeddings}

\begin{figure}[t]
     \centering
     \begin{subfigure}{0.45\columnwidth}
         \includegraphics[width=\textwidth]{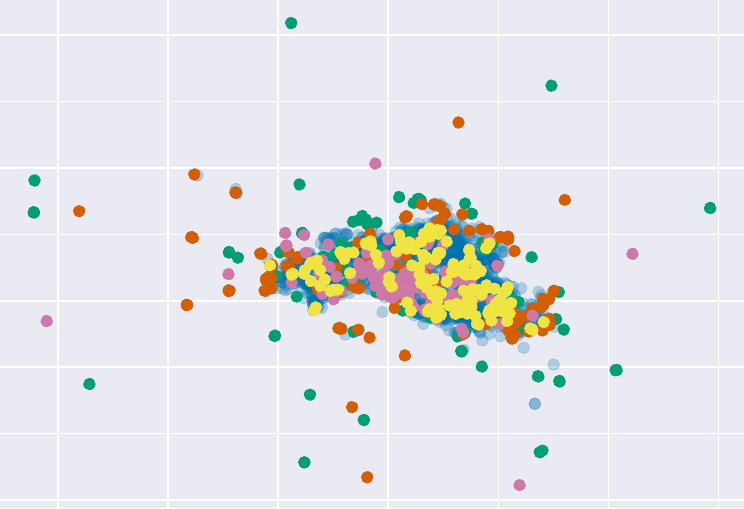}
         \caption{}
         \label{fig:umap_sl}
     \end{subfigure}
     \hfill
     \begin{subfigure}{0.45\columnwidth}
         \includegraphics[width=\textwidth]{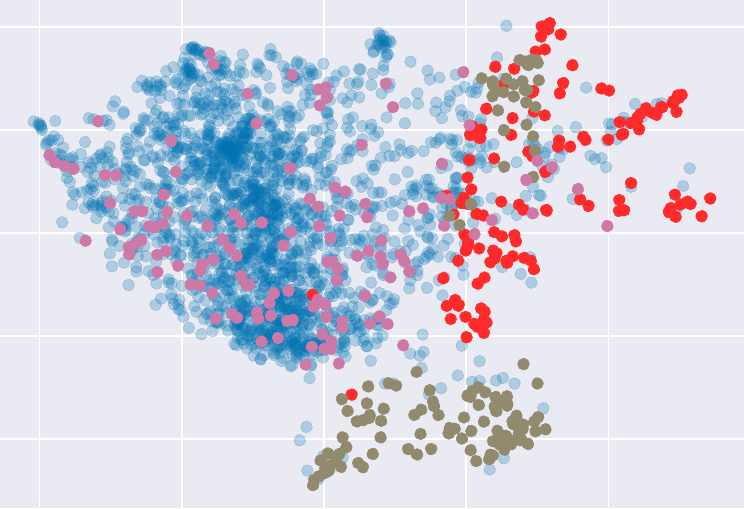}
         \caption{}
         \label{fig:umap_drr}
     \end{subfigure}
     \begin{subfigure}{0.9\columnwidth}
        \centering
         \includegraphics[width=\textwidth]{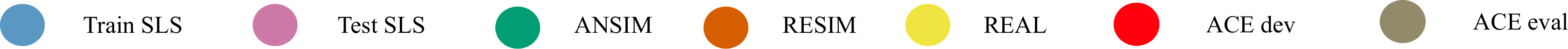}
     \end{subfigure}
    \caption{UMAP projection of different datasets from representations before the \ac{MLP} block of the model. Representations are collected from two networks trained for 3D source localization (a) and DRR (b) using 10,000 samples across all datasets.}
    \label{fig:umap}
    \vspace*{-0.5cm}
\end{figure}

Plotting the representations from our models trained on Spatial LibriSpeech further illustrates the transferability to real-world test data seen in Section~\ref{ssec:transferability}.  Figure~\ref{fig:umap} shows UMAP~\cite{Mcinnes2018} plots of embeddings extracted from representations before the \ac{MLP} block of the network\footnote{UMAP embeddings are constructed by first partitioning the train and test sets into 50 batches of data, then sampling 10,000 samples from all train/test partitions to fit UMAP. Figure~\ref{fig:umap} depicts a random subset of 2048 samples from train Spatial LibriSpeech, and 128 samples from all other evaluation sets.}  We see that the model trained for 3D source localization (Figure \ref{fig:umap_sl}) yields overlapping representations for Spatial LibriSpeech and TUT Sounds Events 2018, while the model trained on DRR (Figure \ref{fig:umap_drr}) yields representations on ACE at the tail of the Spatial LibriSpeech representations indicating the need for a small amount of fine-tuning of the \ac{MLP} block on the target data.

\subsection{Training with a smaller subset of Spatial LibriSpeech}
\label{ssec:lite_experiments}

We explore the performance of models trained with just 10\% of Spatial LibriSpeech\footnote{The number of training epochs is increased by 10x to keep the number of model updates constant.}, which we sample uniformly, maintaining the diversity of the dataset. We find that these models tend to perform worse on the test set of Spatial LibriSpeech than models trained with the full dataset, for instance the best median absolute error on 3D sound localization is 0.39\mdeg higher, and 1.36ms higher for T30 regression on the 10\% models. However, when looking at performance on external baselines, models trained with 10\% of the training data tend to perform  better, for example the best median absolute error on 3D sound localization on TUT Sound Events 2018 was 0.88\mdeg lower, and the median Pearson correlation on T30 regression on ACE Challenge was 0.12 higher. Still, all performance differences were smaller than the IQR between models. Based on these results, we recommend researchers prototype with the smaller version of the dataset, which will be available as a separate download. Additionally, we release the full Spatial LibriSpeech dataset to enable researchers to explore other tasks, such as ablations of acoustic conditions or representation learning.

\section{Conclusion \& Further Work}
\label{sec:conclusion}
We have presented Spatial LibriSpeech, a spatial audio dataset for multiple spatial audio tasks and representation learning. We have shown the utility of our dataset with a simple convolutional network trained on Spatial LibriSpeech, the performance of which transfers to established baselines with minimal intervention, and that the results are close to the state-of-the-art, despite a less sophisticated architecture.

We intend to use Spatial LibriSpeech for a number of other tasks, such as denoising, or room identification. Another interesting line of research is whether we can further improve performance using representation learning to train a single model to regress to many of the spatial audio tasks with the same embedding. We look forward to the community using Spatial LibriSpeech to accelerate research in spatial audio.


\bibliographystyle{IEEEtran}
\bibliography{references.bib}

\begin{thebibliography}{10}
\providecommand{\url}[1]{#1}
\csname url@samestyle\endcsname
\providecommand{\newblock}{\relax}
\providecommand{\bibinfo}[2]{#2}
\providecommand{\BIBentrySTDinterwordspacing}{\spaceskip=0pt\relax}
\providecommand{\BIBentryALTinterwordstretchfactor}{4}
\providecommand{\BIBentryALTinterwordspacing}{\spaceskip=\fontdimen2\font plus
\BIBentryALTinterwordstretchfactor\fontdimen3\font minus
  \fontdimen4\font\relax}
\providecommand{\BIBforeignlanguage}[2]{{%
\expandafter\ifx\csname l@#1\endcsname\relax
\typeout{** WARNING: IEEEtran.bst: No hyphenation pattern has been}%
\typeout{** loaded for the language `#1'. Using the pattern for}%
\typeout{** the default language instead.}%
\else
\language=\csname l@#1\endcsname
\fi
#2}}
\providecommand{\BIBdecl}{\relax}
\BIBdecl

\bibitem{ohuchi2006}
M.~Ohuchi, Y.~Iwaya, Y.~Suzuki, and T.~Munekata, ``A comparative study of sound
  localization acuity of congenital blind and sighted people,'' \emph{Acoust.
  Sci. and Tech.}, vol.~27, no.~5, pp. 290--293, 2006.

\bibitem{hameed2004}
S.~Hameed, H.~Pakarinen, K.~Valde, and V.~Pulkki, ``Psychoacoustic cues in room
  size perception,'' \emph{J. of the Audio Eng. Soc.}, 2004.

\bibitem{LibriSpeech}
V.~Panayotov, G.~Chen, D.~Povey, and S.~Khudanpur, ``{Librispeech: An ASR
  corpus based on public domain audio books},'' in \emph{ICASSP}.\hskip 1em
  plus 0.5em minus 0.4em\relax IEEE, 2015.

\bibitem{MS_DNS}
C.~K. Reddy, H.~Dubey, K.~Koishida, A.~Nair, V.~Gopal, R.~Cutler, S.~Braun,
  H.~Gamper, R.~Aichner, and S.~Srinivasan, ``{INTERSPEECH 2021 Deep Noise
  Suppression Challenge},'' in \emph{Interspeech}, 2021.

\bibitem{Pinardi2021}
D.~Pinardi and A.~Farina, ``Metrics for evaluating the spatial accuracy of
  microphone arrays,'' in \emph{I3DA}.\hskip 1em plus 0.5em minus 0.4em\relax
  IEEE, 2021.

\bibitem{ambisonics}
M.~J. Gerzon, ``Periphone (with height sound reproduction),'' in \emph{Audio
  Eng. Soc. Convention}.\hskip 1em plus 0.5em minus 0.4em\relax AES, 1972.

\bibitem{gamper16}
H.~Gamper, M.~R. Thomas, L.~Corbin, and I.~Tashev, ``{Synthesis of
  Device-Independent Noise Corpora for Realistic ASR Evaluation},'' in
  \emph{Interspeech}, 2016.

\bibitem{TUTSoundEvents}
S.~Adavanne, A.~Politis, J.~Nikunen, and T.~Virtanen, ``Sound event
  localization and detection of overlapping sources using convolutional
  recurrent neural networks,'' \emph{J. of Sel. Top. in Signal Process.},
  vol.~13, no.~1, 2019.

\bibitem{ACE}
J.~Eaton, N.~D. Gaubitch, A.~H. Moore, and P.~A. Naylor, ``{Estimation of Room
  Acoustic Parameters: The ACE Challenge},'' \emph{Trans. on Audio, Speech, and
  Lang. Process.}, vol.~24, 2016.

\bibitem{dEchorate}
D.~Di~Carlo, P.~Tandeitnik, C.~Foy, N.~Bertin, A.~Deleforge, and S.~Gannot,
  ``{dEchorate: a calibrated room impulse response dataset for echo-aware
  signal processing},'' \emph{J. of Audio Speech Music Proc.}, vol.~39, 2021.

\bibitem{Arni}
T.~McKenzie, L.~McCormack, and C.~Hold, ``Dataset of spatial room impulse
  responses in a variable acoustics room for six degrees-of-freedom rendering
  and analysis.''\hskip 1em plus 0.5em minus 0.4em\relax arXiv, 2021.

\bibitem{GIR}
R.~Rust, A.~Xydis, K.~Heutschi, N.~Perraudin, G.~Casas, C.~Du, J.~Strauss,
  K.~Eggenschwiler, F.~Perez-Cruz, F.~Gramazio, and M.~Kohler, ``A data
  acquisition setup for data driven acoustic design,'' \emph{Build. Acoust.},
  vol.~28, no.~4, pp. 345--360, 2021.

\bibitem{EasyCom}
J.~Donley, V.~Tourbabin, J.-S. Lee, M.~Broyles, H.~Jiang, J.~Shen, M.~Pantic,
  V.~K. Ithapu, and R.~Mehra, ``Easycom: An augmented reality dataset to
  support algorithms for easy communication in noisy environments.''\hskip 1em
  plus 0.5em minus 0.4em\relax arXiv, 2021.

\bibitem{CoupledRooms}
T.~McKenzie, S.~J. Schlecht, and V.~Pulkki, ``Acoustic analysis and dataset of
  transitions between coupled rooms,'' in \emph{ICASSP}.\hskip 1em plus 0.5em
  minus 0.4em\relax IEEE, 2021, pp. 481--485.

\bibitem{DCASE}
A.~Politis, S.~Adavanne, D.~Krause, A.~Deleforge, P.~Srivastava, and
  T.~Virtanen, ``A dataset of dynamic reverberant sound scenes with directional
  interferers for sound event localization and detection,'' in \emph{Detect.
  and Classif. of Acoust. Scenes and Events Workshop}, 2021.

\bibitem{ReverbDB}
I.~Szöke, M.~Skácel, L.~Mošner, J.~Paliesek, and J.~Černocký, ``Building
  and evaluation of a real room impulse response dataset,'' \emph{J. of Sel.
  Top. in Signal Process.}, vol.~13, no.~4, 2019.

\bibitem{SBSBRIR}
D.~Satongar, Y.~W. Lam, and C.~Pike, ``Measurement and analysis of a spatially
  sampled binaural room impulse response dataset,'' in \emph{Int. Congr. on
  Sound and Vib.}, 2014.

\bibitem{BIRD}
F.~Grondin, J.-S. Lauzon, S.~Michaud, M.~Ravanelli, and F.~Michaud, ``{BIRD:
  Big Impulse Response Dataset}.''\hskip 1em plus 0.5em minus 0.4em\relax
  arXiv, 2020.

\bibitem{KemarBRIRs}
C.~Mittag, S.~Werner, M.~Böhme, and F.~Klein, ``{Dataset of Binaural Room
  Impulse Responses at Multiple Recording Positions, Source Positions and
  Orientations in a Real Room},'' in \emph{DAGA}.\hskip 1em plus 0.5em minus
  0.4em\relax DEGA Akustik, 2017.

\bibitem{Motus}
G.~Götz, S.~J. Schlecht, and V.~Pulkki, ``A dataset of higher-order ambisonic
  room impulse responses and 3d models measured in a room with varying
  furniture,'' in \emph{I3DA}.\hskip 1em plus 0.5em minus 0.4em\relax IEEE,
  2021.

\bibitem{Aachen}
M.~Jeub, M.~Schafer, and P.~Vary, ``A binaural room impulse response database
  for the evaluation of dereverberation algorithms,'' in \emph{Int. Conf. on
  Digit. Signal Process.}\hskip 1em plus 0.5em minus 0.4em\relax IEEE, 2009.

\bibitem{DIRHA}
M.~Ravanelli, L.~Cristoforetti, R.~Gretter, M.~Pellin, A.~Sosi, and M.~Omologo,
  ``{The DIRHA-ENGLISH corpus and related tasks for distant-speech recognition
  in domestic environments},'' in \emph{ASRU}.\hskip 1em plus 0.5em minus
  0.4em\relax IEEE, 2015.

\bibitem{VoiceHome}
N.~Bertin, E.~Camberlein, E.~Vincent, R.~Lebarbenchon, S.~Peillon,
  {\'E}.~Lamand{\'e}, S.~Sivasankaran, F.~Bimbot, I.~Illina, A.~Tom, S.~Fleury,
  and E.~Jamet, ``{A French corpus for distant-microphone speech processing in
  real homes},'' in \emph{{Interspeech}}, 2016.

\bibitem{SweetHome}
M.~Vacher, B.~Lecouteux, P.~Chahuara, F.~Portet, B.~Meillon, and N.~Bonnefond,
  ``{The Sweet-Home speech and multimodal corpus for home automation
  interaction},'' in \emph{LREC}, 2014.

\bibitem{Grumiaux2022}
P.-A. Grumiaux, S.~Kitić, L.~Girin, and A.~Guérin, ``A survey of sound source
  localization with deep learning methods,'' \emph{J. of the Acoust. Soc. Am.},
  vol. 152, no.~1, 2022.

\bibitem{Gburrek2020}
T.~Gburrek, J.~Schmalenstroeer, A.~Brendel, W.~Kellermann, and R.~Haeb-Umbach,
  ``Deep neural network based distance estimation for geometry calibration in
  acoustic sensor networks,'' in \emph{Eusipco}, 2021.

\bibitem{Ahuja2020}
K.~Ahuja, A.~Kong, M.~Goel, and C.~Harrison, ``{Direction-of-Voice (DoV)
  Estimation for Intuitive Speech Interaction with Smart Devices Ecosystems},''
  in \emph{UIST}.\hskip 1em plus 0.5em minus 0.4em\relax ACM, 2020.

\bibitem{Bryan2020}
N.~J. Bryan, ``Impulse response data augmentation and deep neural networks for
  blind room acoustic parameter estimation,'' in \emph{ICASSP}, 2020.

\bibitem{Pelzer2013}
S.~Pelzer and M.~Vorländer, ``Inversion of a room acoustics model for the
  determination of acoustical surface properties in enclosed spaces,''
  \emph{Proc. of Meet. on Acoust.}, vol.~19, no.~1, 2013.

\bibitem{Srivastava2021}
P.~Srivastava, A.~Deleforge, and E.~Vincent, ``Blind room parameter estimation
  using multiple multichannel speech recordings,'' in \emph{Workshop on Appl.
  of Signal Process. to Audio and Acoust.}, 2021.

\bibitem{Kinoshita2017}
K.~Kinoshita, M.~Delcroix, H.~Kwon, T.~Mori, and T.~Nakatani, ``Neural
  network-based spectrum estimation for online wpe dereverberation.'' in
  \emph{Interspeech}, 2017.

\bibitem{Heymann2016}
J.~Heymann, L.~Drude, and R.~Haeb-Umbach, ``Neural network based spectral mask
  estimation for acoustic beamforming,'' in \emph{ICASSP}, 2016.

\bibitem{Tzinis2020}
E.~Tzinis, Z.~Wang, and P.~Smaragdis, ``Sudo rm -rf: Efficient networks for
  universal audio source separation,'' in \emph{Int. Workshop on Mach. Learn.
  for Signal Process.}\hskip 1em plus 0.5em minus 0.4em\relax IEEE, 2020.

\bibitem{Gong2019}
Z.~Gong, P.~Zhong, and W.~Hu, ``Diversity in machine learning,'' \emph{IEEE
  Access}, vol.~7, 2019.

\bibitem{data2vec}
A.~Baevski, W.-N. Hsu, Q.~Xu, A.~Babu, J.~Gu, and M.~Auli, ``data2vec: A
  general framework for self-supervised learning in speech, vision and
  language,'' in \emph{ICML}, 2022.

\bibitem{Diaz2005}
C.~Díaz and A.~Pedrero, ``The reverberation time of furnished rooms in
  dwellings,'' \emph{Appl. Acoustics}, vol.~66, no.~8, 2005.

\bibitem{ITU-T-P.581}
{ITU-T P.581}, ``{Use of head and torso simulator for hands-free and handset
  terminal testing},'' International Telecommunication Union, {standard}, Jul.
  2022.

\bibitem{ISO-3382-1}
{ISO 3382-1}, ``{Measurement of room acoustic parameters — Part 1:
  Performance spaces},'' International Organization for Standardization,
  standard, Jun. 2009.

\bibitem{pilot2021}
C.~Schymura, B.~Bönninghoff, T.~Ochiai, M.~Delcroix, K.~Kinoshita,
  T.~Nakatani, S.~Araki, and D.~Kolossa, ``{PILOT: Introducing Transformers for
  Probabilistic Sound Event Localization},'' in \emph{Interspeech}, 2021.

\bibitem{Jacobsen1990}
F.~Jacobsen, ``Active and reactive sound intensity in a reverberant sound
  field,'' \emph{J of Sound and Vib.}, vol. 143, no.~2, 1990.

\bibitem{rafaely2015}
B.~Rafaely, \emph{Fundamentals of spherical array processing}.\hskip 1em plus
  0.5em minus 0.4em\relax Springer, 2015, vol.~8.

\bibitem{Mcinnes2018}
L.~McInnes, J.~Healy, and J.~Melville, ``{UMAP: Uniform manifold approximation
  and projection for dimension reduction}.''\hskip 1em plus 0.5em minus
  0.4em\relax arXiv, 2018.

\end{thebibliography}

\end{document}